\pdfoutput=1  

\documentclass[aps,twocolumn,groupedaddress,floatfix,preprintnumbers,nofootinbib,eqsecnum]{revtex4}

\usepackage{amsmath, float, graphicx,placeins,amssymb,multirow,pgfplots,pgfplotstable}
\usepackage{tikz,hyperref}
\usepackage{CJKutf8}
\usetikzlibrary{shapes.geometric,arrows}

\begin{document}
\title{
Constraining Dark-Matter Ensembles with Supernova Data  
}

\author{Aditi Desai$^{1}$\footnote{E-mail address:  {\tt desaia@lafayette.edu}},
  Keith R.\ Dienes$^{2,3}$\footnote{E-mail address:  {\tt dienes@email.arizona.edu}},
  Brooks Thomas$^{1}$\footnote{E-mail address:  {\tt thomasbd@lafayette.edu}}}
\affiliation{
  $^1\,$Department of Physics, Lafayette College, Easton, PA  18042  USA\\
  $^2\,$Department of Physics, University of Arizona, Tucson, AZ  85721  USA\\
  $^3\,$Department of Physics, University of Maryland, College Park, MD 20742 USA}

\begin{abstract}
The constraints on non-minimal dark sectors involving ensembles of unstable 
dark-matter species are well established and quite stringent in cases in which
these species decay to visible-sector particles.  However, in cases in which
these ensembles decay exclusively to other, lighter dark-sector states, the 
corresponding constraints are less well established.  In this paper, we investigate 
how information about the expansion rate of the universe at low redshifts gleaned 
from observations of Type~Ia supernovae can be used to constrain ensembles of unstable 
particles which decay primarily into dark radiation.
\end{abstract}

\maketitle


\newcommand{\PRE}[1]{{#1}} 
\newcommand{\ul}{\underline}
\newcommand{\del}{\partial}
\newcommand{\nbox}{{\,\lower0.9pt\vbox{\hrule \hbox{\vrule height 0.2 cm
\hskip 0.2 cm \vrule height 0.2 cm}\hrule}\,}}

\newcommand{\postscript}[2]{\setlength{\epsfxsize}{#2\hsize}
   \centerline{\epsfbox{#1}}}
\newcommand{\gweak}{g_{\text{weak}}}
\newcommand{\mweak}{m_{\text{weak}}}
\newcommand{\mplanck}{M_{\text{Pl}}}
\newcommand{\mstar}{M_{*}}
\newcommand{\sigmaan}{\sigma_{\text{an}}}
\newcommand{\sigmatot}{\sigma_{\text{tot}}}
\newcommand{\sigmaSI}{\sigma_{\rm SI}}
\newcommand{\sigmaSD}{\sigma_{\rm SD}}
\newcommand{\OmegaM}{\Omega_{\text{M}}}
\newcommand{\OmegaDM}{\Omega_{\text{DM}}}
\newcommand{\ipb}{\text{pb}^{-1}}
\newcommand{\ifb}{\text{fb}^{-1}}
\newcommand{\iab}{\text{ab}^{-1}}
\newcommand{\ev}{\text{eV}}
\newcommand{\kev}{\text{keV}}
\newcommand{\mev}{\text{MeV}}
\newcommand{\gev}{\text{GeV}}
\newcommand{\tev}{\text{TeV}}
\newcommand{\pb}{\text{pb}}
\newcommand{\mb}{\text{mb}}
\newcommand{\cm}{\text{cm}}
\newcommand{\m}{\text{m}}
\newcommand{\km}{\text{km}}
\newcommand{\kg}{\text{kg}}
\newcommand{\g}{\text{g}}
\newcommand{\s}{\text{s}}
\newcommand{\yr}{\text{yr}}
\newcommand{\Mpc}{\text{Mpc}}
\newcommand{\etal}{{\em et al.}}
\newcommand{\eg}{{\em e.g.}}
\newcommand{\ie}{{\em i.e.}}
\newcommand{\ibid}{{\em ibid.}}
\newcommand{\Eqref}[1]{Equation~(\ref{#1})}
\newcommand{\secref}[1]{Sec.~\ref{sec:#1}}
\newcommand{\secsref}[2]{Secs.~\ref{sec:#1} and \ref{sec:#2}}
\newcommand{\Secref}[1]{Section~\ref{sec:#1}}
\newcommand{\appref}[1]{App.~\ref{sec:#1}}
\newcommand{\figref}[1]{Fig.~\ref{fig:#1}}
\newcommand{\figsref}[2]{Figs.~\ref{fig:#1} and \ref{fig:#2}}
\newcommand{\Figref}[1]{Figure~\ref{fig:#1}}
\newcommand{\tableref}[1]{Table~\ref{table:#1}}
\newcommand{\tablesref}[2]{Tables~\ref{table:#1} and \ref{table:#2}}
\newcommand{\Dsle}[1]{\slash\hskip -0.28 cm #1}
\newcommand{\met}{{\Dsle E_T}}
\newcommand{\mpt}{\not{\! p_T}}
\newcommand{\Dslp}[1]{\slash\hskip -0.23 cm #1}
\newcommand{\Dsl}[1]{\slash\hskip -0.20 cm #1}

\newcommand{\mB}{m_{B^1}}
\newcommand{\mq}{m_{q^1}}
\newcommand{\mf}{m_{f^1}}
\newcommand{\mKK}{m_{KK}}
\newcommand{\WIMP}{\text{WIMP}}
\newcommand{\SWIMP}{\text{SWIMP}}
\newcommand{\NLSP}{\text{NLSP}}
\newcommand{\LSP}{\text{LSP}}
\newcommand{\mWIMP}{m_{\WIMP}}
\newcommand{\mSWIMP}{m_{\SWIMP}}
\newcommand{\mNLSP}{m_{\NLSP}}
\newcommand{\mchi}{m_{\chi}}
\newcommand{\mgravitino}{m_{\gravitino}}
\newcommand{\mmed}{M_{\text{med}}}
\newcommand{\gravitino}{\tilde{G}}
\newcommand{\Bino}{\tilde{B}}
\newcommand{\photino}{\tilde{\gamma}}
\newcommand{\stau}{\tilde{\tau}}
\newcommand{\slepton}{\tilde{l}}
\newcommand{\snu}{\tilde{\nu}}
\newcommand{\squark}{\tilde{q}}
\newcommand{\mgaugino}{M_{1/2}}
\newcommand{\epsEM}{\varepsilon_{\text{EM}}}
\newcommand{\mmess}{M_{\text{mess}}}
\newcommand{\lmess}{\Lambda}
\newcommand{\nmess}{N_{\text{m}}}
\newcommand{\signmu}{\text{sign}(\mu)}
\newcommand{\Omegachi}{\Omega_{\chi}}
\newcommand{\lambdafs}{\lambda_{\text{FS}}}
\newcommand{\be}{\begin{equation}}
\newcommand{\ee}{\end{equation}}
\newcommand{\bea}{\begin{eqnarray}}
\newcommand{\eea}{\end{eqnarray}}
\newcommand{\beq}{\begin{equation}}
\newcommand{\eeq}{\end{equation}}
\newcommand{\beqn}{\begin{eqnarray}}
\newcommand{\eeqn}{\end{eqnarray}}
\newcommand{\baln}{\begin{align}}
\newcommand{\ealn}{\end{align}}
\newcommand{\lsim}{\lower.7ex\hbox{$\;\stackrel{\textstyle<}{\sim}\;$}}
\newcommand{\gsim}{\lower.7ex\hbox{$\;\stackrel{\textstyle>}{\sim}\;$}}

\newcommand{\ssection}[1]{{\em #1.\ }}
\newcommand{\rem}[1]{\textbf{#1}}

\def\ie{{\it i.e.}\/}
\def\eg{{\it e.g.}\/}
\def\etc{{\it etc}.\/}
\def\calN{{\cal N}}

\def\mptwo{{m_{\pi^0}^2}}
\def\mp{{m_{\pi^0}}}
\def\sqtsn{\sqrt{s_n}}
\def\sqtsn{\sqrt{s_n}}
\def\sqtsn{\sqrt{s_n}}
\def\sqts0{\sqrt{s_0}}
\def\Dsqts{\Delta(\sqrt{s})}
\def\Omegatot{\Omega_{\mathrm{tot}}}
\def\rhotot{\rho_{\mathrm{tot}}}
\def\rhocrit{\rho_{\mathrm{crit}}}
\def\OmegaDM{\Omega_{\mathrm{DM}}}
\def\OmegaDMbar{\overline{\Omega}_{\mathrm{DM}}}
\def\tLS{t_{\mathrm{LS}}}
\def\aLS{a_{\mathrm{LS}}}
\def\zLS{z_{\mathrm{LS}}}
\def\tnow{t_{\mathrm{now}}}
\def\znow{z_{\mathrm{now}}}
\def\Ndof{N_{\mathrm{dof}}\/}
\def\tprod{t_{\mathrm{prod}}}


\section{Introduction\label{sec:Intro}}


Dynamical Dark Matter (DDM)~\cite{DDM1,DDM2} is an alternative framework for dark-matter 
physics in which the dark matter is an ensemble comprising 
a potentially vast number of different constituent particle species whose properties 
(masses, lifetimes, cosmological abundances, \etc) scale across the ensemble 
according to a set of scaling relations.  The specific scaling relations depend on 
the underlying physics of the particular DDM model in question.  In all cases, however, 
these scaling relations lead to a balancing of decay widths against cosmological 
abundances across the ensemble such that the abundances of the more unstable
constituents are suppressed.  In this way, the DDM framework
circumvents the stringent bounds on dark-matter decays in traditional
dark-matter scenarios --- scenarios in which the dark-matter candidate $\chi$ has a 
single, well-defined lifetime $\tau_\chi$.  

Observational constraints on dark-matter decay --- together with the traditional 
assumption that $\chi$ contributes essentially the entirety 
of the total present-day dark-matter abundance --- impose a stringent lower bound on $\tau_\chi$.  
Indeed, in such scenarios, one finds that $\chi$ must be ``hyperstable,'' with a 
lifetime which significantly exceeds the present age of the universe.  By contrast, within 
the DDM framework, a non-trivial fraction of the total abundance of 
the dark-matter ensemble can be carried by particle species with lifetimes well below the
timescale associated with this hyperstability bound without violating the same observational 
constraints.  In this way, the DDM framework evades the stringent 
bounds that arise for dark-matter decays in traditional dark-matter scenarios, thereby 
broadening the theory space of viable decaying-dark-matter models.  Moreover, realizations
of this framework can give rise to novel signatures at 
colliders~\cite{DDMLHC,CutsAndCorrelations}, at dedicated long-lived-particle 
detectors~\cite{MATHUSLAWhitePaper,DDMMATHUSLA}, at direct-detection 
experiments~\cite{DDMDD}, and at indirect-detection 
experiments~\cite{DDMAMS,DDMBoxLine1,DDMBoxLine2}.

The hyperstability bound on the lifetime of a traditional dark-matter candidate 
depends crucially on the final states into which it decays.  
In scenarios in which $\chi$ decays with a non-negligible branching fraction into final 
states involving any Standard-Model (SM) particles other than neutrinos, constraints on the 
diffuse photon flux from Fermi-LAT~\cite{Fermi} imply a hyperstability bound of   
$\tau_\chi \gtrsim \mathcal{O}(10^{28}\mathrm{~s})$ for dark-matter masses in the range 
$\mathcal{O}(100\mathrm{~MeV}) \lesssim m_\chi 
\lesssim \mathcal{O}(\mathrm{EeV})$~\cite{HyperstabilityCohen,HyperstabilityCirelli,
HyperstabilityLiu,Hooperstability}.  The corresponding constraints for a dark-matter
mass in the range $\mathcal{O}(10\mathrm{~keV}) \lesssim m_\chi 
\lesssim \mathcal{O}(100\mathrm{~MeV})$ from measurements of the diffuse photon flux
at lower energies~\cite{EssigLightDMHyperstability} and from CMB 
data~\cite{SlatyerLightDMCMBBound}, while slightly more dependent on the 
channels through which $\chi$ decays, are nevertheless quite stringent.  
Measurements of the positron flux by the AMS-02 detector~\cite{AMSFirstResults,AMSUpdate}  
also imply constraints of roughly the same order on dark-matter decays to a wide 
variety of final states~\cite{IbarraAMS}.    

By contrast, if the dark-matter candidate decays exclusively to other, lighter states 
within the dark sector, the hyperstability bound on its lifetime is far weaker.
The leading constraints on dark-matter decays of this sort ultimately stem from the fact
that the conversion of the mass energy of the decaying dark-matter particles into the 
kinetic energy of their decay products alters the effective equation of state of the dark 
sector as a whole.  This in turn leads a modification of the expansion history of the 
universe relative to the prediction of the standard cosmology.  Such a modification would not 
only leave imprints both in the power spectrum of the CMB and in the matter power spectrum, 
but would also be evident in baryon acoustic oscillations (BAO) and in Type~Ia supernova data, 
both of which provide an observational handle on the expansion history at low redshifts.  
Taken together, these considerations place the hyperstability bound on the lifetime of a 
traditional dark-matter candidate which decays exclusively to other states within the dark 
sector at around 
$\tau_\chi \gtrsim 2 \times 10^{19}$~s~\cite{GongChenDarkBound,PoulinDarkToDarkBound}.

Within explicit realizations of the DDM framework, such as those described in
Refs.~\cite{DDMAxion,DDMHagedorn,DDMThermal}, the situation can be different.  
Indeed, much effort has already been devoted to determining the extent to which the 
hyperstability bound on dark-matter decays into visible-sector particles can be 
circumvented in such cases. 
However, the extent to which the hyperstability bound on dark-matter decays solely to 
other particles within the dark sector can be circumvented within this framework has yet 
to be explored.  In this paper, we take a first step in this direction.  
In particular, we investigate how information about the expansion rate of the
universe at low redshifts gleaned from the relationship between the redshifts and 
luminosity distances of Type~Ia supernovae can be used to constrain DDM ensembles
which decay primarily to other, lighter states within the dark sector which act as
dark radiation.  This technique for constraining decays within the dark sector, which 
has previously been applied to scenarios involving a single unstable particle 
species~\cite{BlackadderKoushiappas1,BlackadderKoushiappas2}, is particularly relevant 
for constraining scenarios within the DDM framework.
  
The outline of this paper is as follows.  In Sect.~\ref{sec:DDMEnsemble}, we 
describe the DDM model on which we shall focus in this paper.  As we shall see,
this model is representative of a large class of models within the DDM framework.
In Sect.~\ref{sec:Cosmology}, we derive an expression for the luminosity distance 
$d_L(z)$ as a function of cosmological redshift $z$ for an ensemble of unstable
particles which decay to other, lighter states within the dark sector.
In Sect.~\ref{sec:SNData}, we describe the catalog of Type~Ia supernovae we 
use in order to constrain the relationship between redshift and luminosity distance in 
scenarios involving dark-to-dark decays.  We also outline our statistical method
for assessing the goodness of fit between the functional form of $d_L(z)$ obtained
within any such scenario and the set of measured redshifts and luminosity distances 
for the supernovae within this data set.  Finally, in Sect.~\ref{sec:Results}, we present 
our results and assess the extent to which our DDM parameter space can 
be constrained by supernova data.  In Sect.~\ref{sec:Conclusions}, we conclude with 
a summary of our findings and a discussion of possible avenues for future work.


\section{Parametrizing the DDM Ensemble\label{sec:DDMEnsemble}}


Within the DDM framework, the dark-matter candidate is an ensemble comprising
a large number $N$ of individual constituent particle species $\chi_n$, where the
index $n=0,1,2,\ldots,N-1$ labels the particles in order of increasing mass. 
In many realizations of DDM, the spectrum of masses $m_n$ for these ensemble constituents
takes the form
\begin{equation}
  m_n ~=~ m_0 + n^\delta \Delta m~,
\end{equation}
where $m_0$ denotes the mass of the lightest ensemble constituent, where $\Delta m$ is 
a free parameter with dimensions of mass, and where $\delta$ is a dimensionless 
scaling exponent.  

For simplicity, we shall assume that each ensemble constituent decays with a
branching fraction of essentially unity via the process 
$\chi_n \rightarrow \overline{\psi}\psi$, where $\psi$ is a massless dark-sector
particle which behaves as dark radiation.  Moreover, we shall also assume that the 
total decay widths $\Gamma_n$ of the $\chi_n$ scale across the ensemble according to a 
power law of the form 
\begin{equation}
  \Gamma_n ~=~ \Gamma_0 \left(\frac{m_n}{m_0}\right)^\xi~,
\end{equation}
where $\Gamma_0$ denotes the decay width of the lightest 
particle in the ensemble and where $\xi$ is another dimensionless scaling exponent.
In what follows, we take $\Gamma_0$ and $\xi$ to be free parameters.

We note that intra-ensemble decays --- \ie, decays of the ensemble constituents 
into final states which include other, lighter $\chi_n$ --- represent another
important class of dark-to-dark decays which can potentially occur within the
DDM framework.  Indeed, such decays arise in many realizations of the DDM framework.  
While we shall not consider intra-ensemble decays in this analysis, we note that 
supernova constraints on scenarios in which one or more of the ensemble constituents 
have non-negligible branching fractions into final states involving other, lighter 
$\chi_n$ are generically weaker than the bounds on scenarios in which the ensemble 
constituents decay to states involving dark radiation alone. 
This is because these constraints ultimately follow from deviations in the expansion
rate of the universe as a function of redshift from the expected relationship
obtained within a $\Lambda_{\mathrm{CDM}}$ cosmology.  Thus, the constraints 
we shall derive in this paper for a given DDM model represent a conservative 
bound on extensions of this same model involving intra-ensemble decays.   

We shall assume that the initial abundances $\Omega_n$ for the individual ensemble 
constituents are established at some early time $\tprod$.  We shall assume that 
$\tprod \ll \tLS \approx 1.17 \times 10^{13}$~s, where $\tLS$ denotes the time 
of last scattering, and that $\tprod \ll \tau_{N-1}$.  However, provided that these 
two criteria are satisfied, our results in what follows will be independent of $\tprod$.  
In order to retain as much generality as possible, we  
shall remain largely agnostic about the mechanism which 
generates these abundances.  However, we shall assume that the cosmological 
population of each ensemble constituent can be considered to be ``cold,''
in the sense that its equation-of-state parameter may be taken to be $w_n \approx 0$ 
for all $t > \tprod$.  Moreover, for concreteness, we shall 
assume that the initial abundances $\Omega_n(\tprod)$ of the individual ensemble 
constituents at $t = t_{\rm prod}$ scale across the ensemble according to a power law of the 
form     
\begin{equation}
  \Omega_n(\tprod) ~=~ \Omega_0(\tprod)\left(\frac{m_n}{m_0}\right)^\gamma~,
  \label{eq:AbundanceScaling}
\end{equation} 
where $\Omega_0(\tprod)$ denotes the initial abundance of the lightest ensemble constituent 
and where $\gamma$ is a dimensionless scaling exponent.  We take this scaling exponent to be 
a free parameter.  By contrast, as we shall discuss in further detail in Sect.~\ref{sec:SNData}, 
the value of $\Omega_0(\tprod)$ is essentially fixed by the requirement that the total initial 
abundance $\Omega_{\mathrm{tot}}(\tprod) ~\equiv~ \sum_n \Omega_n(\tprod)$ of the DDM ensemble at 
$t = \tLS$ accord with the dark-matter abundance $\OmegaDM(\tLS)$ 
derived from Planck data~\cite{Planck}.


\section{Cosmic Expansion in the Presence of Decaying Ensembles\label{sec:Cosmology}}


Observational data~\cite{Planck} indicate that at large scales our universe is 
extremely homogeneous, isotropic, and spatially flat.  Such a universe is 
described by a Friedmann-Robertson-Walker (FRW) metric with vanishing spatial curvature.  
The expansion rate of the universe in an FRW cosmology may be quantified in terms of the 
Hubble parameter  $H = \dot{a}/a$, where $a$ is the scale factor.  In such a 
universe, the luminosity distance $d_L(z)$ of an astrophysical source, expressed as a 
function of its cosmological redshift $z \equiv (1-a)/a$, is    
\begin{equation}
  d_L(z) ~=~ \frac{c(1+z)}{H_{\mathrm{now}}} \int_0^z \mathcal{F}^{-1/2}(z')dz'~,
  \label{wq:LuminosityDistance}
\end{equation}
where $H_{\mathrm{now}}$ is the present-day value of the Hubble parameter and where 
the quantity 
\begin{equation}
  \mathcal{F}(z') ~\equiv~ \frac{1}{\rhocrit(0)}\sum_i \rho_i(z')
\end{equation}
represents the sum over the energy densities of all relevant cosmological 
components (photons, baryons, dark matter, \etc), normalized to the value of 
the critical energy density $\rhocrit(0)$ at redshift $z = 0$ --- \ie, at present 
time.  More specifically, for the toy DDM model defined in Sect.~\ref{sec:DDMEnsemble},
we have
\begin{eqnarray}
  \mathcal{F}(z) &=& \frac{1}{\rhocrit(0)}\Big[   
    \rhotot(z) + \rho_\psi(z) + 
    \rho_b(z) \nonumber \\ & & ~~~+ \rho_\gamma(z) + 
    \rho_\nu(z) + \rho_\Lambda(z) \Big]
  \label{eq:DefOfF}
\end{eqnarray}
where $\rho_{\mathrm{tot}}(z)$, $\rho_\psi(z)$, $\rho_b(z)$, $\rho_\gamma(z)$, 
$\rho_\nu(z)$, and $\rho_\Lambda(z)$ respectively denote the energy densities 
of the DDM ensemble as a whole, the dark-radiation field $\psi$, baryons, photons, 
neutrinos, and dark energy.  Thus, in order to determine the functional relationship 
between the redshifts and luminosity distances of astrophysical objects for any
given choice of parameters within this model, we must assess how each of these 
energy densities evolves as a function of $z$.

In general, the energy density of a cosmological component with equation-of-state
parameter $w_i(z)$ scales with $z$ according to the relation
\begin{equation}
  \rho_i(z) ~=~ \rho_i(0)\, (1+z)^{3[1+w_i(z)]}~,
\end{equation}
where $\rho_i(0)$ denotes the energy density of that component at present time.
For those cosmological components for which $w_i(z)$ is effectively constant since
the time of last scattering, $\rho_i(z)$ is trivial 
to obtain.  For example, since $w_b \approx 0$ and $w_\gamma = 1/3$, we have 
$\rho_b(z) ~=~ \rho_b(0) (1+z)^3$ and $\rho_\gamma(z) ~=~ \rho_\gamma(0) (1+z)^{4}$.  
Likewise, under the assumption that the dark energy behaves like a cosmological 
constant --- \ie, that $w_\Lambda \approx -1$ at all times subsequent to $\tLS$ --- we have 
$\rho_\Lambda(z) ~\approx~ \rho_\Lambda(0)$ for all $z$.  The present-day energy densities 
of these cosmological components can be inferred from Planck data~\cite{Planck}.  In 
particular, we find that $\rho_b(0) \approx 2.32\times 10^{-7}\mathrm{~GeV}\,\mathrm{cm}^{-3}$ 
and $\rho_\Lambda(0) \approx 3.24\times 10^{-6}\mathrm{~GeV}\,\mathrm{cm}^{-3}$, while
the energy density of photons at $z=0$ is given by 
\begin{equation}
  \rho_\gamma(0) ~=~ \frac{4\sigma}{c} T_{\gamma}^4(0)~,
\end{equation}
where $\sigma \approx 0.0354\mathrm{~GeV}\,\mathrm{cm}^{-2}\,\mathrm{s}^{-1}\,\mathrm{K}^{-4}$ 
is the Stefan-Boltzmann constant, where $c$ is the speed of light, and where 
$T_{\gamma}(0) \approx 2.73$~K is the present-day CMB-photon temperature.   

The evolution of $\rho_\nu(z)$ with $z$ is slightly more complicated due to the
presence of small but non-zero masses $m_{\nu_i}$ for at least two of the three 
neutrino mass eigenstates.  At early times, all neutrinos species are highly 
relativistic.  At such times, $w_\nu(z) \approx 1/3$ and $\rho_\nu(z)$ scales with 
redshift as $\rho_\nu \propto (1+z)^4$.  Thus, at such times, $\rho_\nu(z)$ is 
proportional to $\rho_\gamma(z)$.  However, as $t$ increases and the temperature
$T_\nu$ in the neutrino sector drops, $\rho_\nu(z)$ eventually begins
to deviate significantly from this simple scaling behavior.  Indeed, under the 
assumption that $m_{\nu_i} > 0$ for all three neutrino species, one would expect 
that $\rho_\nu \propto (1+z)^3$ at sufficiently late times.  In order to 
interpolate between the early-time and late-time behavior of $\rho_\nu(z)$, it is 
traditional to introduce a scaling function $f(z)$ such that
\begin{equation}
  \rho_\nu(z) ~=~ \frac{7}{8} \left(\frac{4}{11}\right)^{4/3} N_{\mathrm{eff}}\,
    \rho_\gamma(z)\, f(z)~,
\end{equation}
where $N_{\mathrm{eff}} \approx 3.042$~\cite{Mangano} is the effective number of 
neutrino species.  The functional form of $f(z)$ turns out to be well approximated 
by~\cite{WMAP7Year} 
\begin{equation}
  f(z) ~\approx~ \left[1 + \left(\frac{A}{1+z}\right)^p\right]^{1/p}~,
\end{equation}
where $p \approx 1.83$ and where the dimensionless constant $A$ is given by 
\begin{equation}
  A ~\approx~ \left(1.87 \times 10^5\mbox{~eV}^{-1}\right) 
    \left[\frac{180\,\zeta(3)}{7\pi^4}\right] 
    \sum_{i=1}^3 m_{\nu_i}~.
  \label{eq:ACoeffinfOfz}
\end{equation}
While the individual neutrino masses $m_{\nu_i}$ are not currently known, the sum 
appearing in Eq.~(\ref{eq:ACoeffinfOfz}) is bounded from above by cosmological 
considerations and from below by neutrino-oscillation data.
Bounds in the literature differ slightly depending on the 
particulars of the analysis method and on whether a normal or inverted neutrino-mass 
hierarchy is assumed.  However, as discussed in Ref.~\cite{VagnozziNeutrinoMassBounds} 
and references therein, this sum is constrained to lie within the rough range
\begin{equation}
  0.06\mbox{~eV} ~\lesssim~ \sum_{i=1}^3 m_{\nu_i} ~\lesssim~ 0.15\mbox{~eV}
\end{equation}
within the context of a $\Lambda$CDM cosmology.  For concreteness, we shall adopt 
$\sum_{i=1}^3 m_{\nu_i} = 0.1$~eV as our benchmark in what follows. 

Finally, we consider the energy densities $\rho_n(z)$ and $\rho_\psi(z)$ of 
the individual ensemble constituents $\chi_n$ and the dark-radiation field $\psi$. 
These energy densities evolve according system of Boltzmann 
equations given by
\begin{eqnarray}
  \frac{d\rho_n}{dt} + 3H\rho_n &=& -\Gamma_n \rho_n \nonumber \\
  \frac{d\rho_\psi}{dt} + 4H\rho_\psi &=& \sum_{n=0}^{N-1} \Gamma_n \rho_n~,
  \label{eq:BoltzmannEqs}
\end{eqnarray}
where collision terms associated with inverse-decay processes of the form 
$\psi\overline{\psi}\rightarrow\chi_n$ have been omitted, as their effect on 
the $\rho_n$ and on $\rho_\psi$ is negligible.  The evolution equation for 
$\rho_n(t)$ may also be expressed in the equivalent form
\begin{equation}
  \frac{d}{dt}\left(a^3\rho_n\right) ~=~ -\Gamma_n (a^3\rho_n)~,
\end{equation}
which may be integrated directly in order to obtain an expression for $\rho_n$ 
as a function of time, or equivalently as a function of the scale factor $a$.
In particular, the expression for $\rho_n(a)$ is found to be 
\begin{equation}
  \rho_n(a) ~=~ \rho_n(a_{\mathrm{LS}})\left(\frac{\aLS}{a}\right)^3
    e^{-\Gamma_n [t(a) - \tLS]}~,
  \label{eq:rhonofaraw}
\end{equation}
where $\aLS$ is the scale factor at last scattering and where $t(a)$ is 
the time in the background frame expressed as a function of $a$. 

In order to solve the Boltzmann equation for $\rho_\psi$ in Eq.~(\ref{eq:BoltzmannEqs}),
we begin by changing variables from $t$ to $a$, yielding 
\begin{equation}
  \frac{d\rho_\psi}{da} ~=~ -\frac{4\rho_\psi}{a} + \frac{1}{aH} 
    \sum_{n=0}^{N-1} \Gamma_n \rho_n(a)~.
\end{equation}
Substituting for $\rho_n(a)$ using Eq.~(\ref{eq:rhonofaraw}), we have
\begin{eqnarray}
  \frac{d\rho_\psi}{da} &=& -\frac{4\rho_\psi}{a} + 
    \frac{\aLS^3\rhocrit(\aLS)\mathcal{F}^{-1/2}(a)}{a^4 H_{\mathrm{now}}} \nonumber \\ 
  & &~~  \times \sum_{n=0}^{N-1} \Gamma_n \Omega_n(\tLS)
    e^{-\Gamma_n [t(a) - \tLS]}~,
  \label{eq:RhoPsiEvolEqSubbed}
\end{eqnarray}
where $\mathcal{F}(a)$ is given by Eq.~(\ref{eq:DefOfF}).
We emphasize that not only does the right side of Eq.~(\ref{eq:RhoPsiEvolEqSubbed})
explicitly involve $t(a)$, but it also implicitly involves both $t(a)$ and 
$\rho_\psi(a)$ through $\mathcal{F}(a)$.  The form of $t(a)$ in our DDM 
scenario may be inferred from the relation   
\begin{equation}
  \frac{dt}{da} ~=~ \frac{1}{Ha} ~=~ 
    \frac{\mathcal{F}^{-1/2}(a)}{a H_{\mathrm{now}}}~,
  \label{eq:dtda}    
\end{equation}
the right side of which likewise involves both $t(a)$ and $\rho_\psi(a)$ 
through $\mathcal{F}(a)$.  Eqs.~(\ref{eq:dtda}) and~(\ref{eq:dtda})  
may therefore be solved together numerically as a coupled system in order to yield 
expressions for $\rho_\psi(a)$ and $t(a)$ as functions of $a$, or equivalently 
as functions of the redshift $z$.  Once these expressions are known, they may be 
substituted into Eq.~(\ref{wq:LuminosityDistance}) in order to obtain a functional 
form for $d_L(z)$.


\section{Implementing Supernova Constraints on Decays within the Dark Sector\label{sec:SNData}}


A constraint on the functional form of $d_L(z)$ within the recent cosmological
past --- \ie, at redshifts $0 < z \lesssim 5$ --- can be derived from observations of the 
redshifts and luminosity distances of Type~Ia supernovae.  The luminosity distance $d_L$ of an 
astrophysical source can be inferred from its distance modulus $\mu$, which represents the 
difference between its apparent and absolute magnitude.  In particular, the relationship 
between these two quantities is given by 
\begin{equation}
  \mu ~=~ 5 \log_{10} \left(\frac{d_{L}}{\mathrm{Mpc}}\right) + 25~.
\end{equation} 
In this analysis, we derive our constraints on $d_L(z)$ from the combined Pantheon 
sample~\cite{Pantheon}, which contains magnitude and redshift information
for $N_{\mathrm{SN}} = 1048$ spectroscopically confirmed Type~Ia supernovae with 
high-quality light curves.

In order to compare the theoretical relationship between $d_L(z)$ and $z$ obtained
for a given choice of our DDM model parameters to the results obtained for the 
Pantheon data, we proceed as follows.  We evaluate the 
goodness-of-fit statistic
\begin{equation}
  \chi^2 ~=~ \sum_{j=1}^{N_{\mathrm{SN}}} \frac{\big[\mu_{j}^\mathrm{obs}
    - \mu(z_j)\big]^2}{(\Delta \mu_j^\mathrm{obs})^2}~,
  \label{eq:ChiSq}
\end{equation}
where the index $j = 1, 2, \ldots, N_{\mathrm{SN}}$ labels the supernovae in the 
data set, where $\mu_{j}^\mathrm{obs}$ represents the observed value of the distance 
modulus for the $j$th supernova in that set, where $\Delta \mu_j^\mathrm{obs}$ is
the uncertainty in $\mu_{j}^\mathrm{obs}$, and where $\mu(z_j)$ is 
the predicted value of the distance modulus obtained for the measured redshift $z_j$ of 
that same supernova within the context of a particular cosmological model.
In order to assess how the supernovae in the Pantheon sample constrain the parameter
space of our toy DDM model, we proceed as follows.  We first obtain a $p$-value by 
comparing the value of $\chi^2$ obtained for any particular choice of model parameters 
to a chi-square distribution with $\Ndof = N_{\mathrm{SN}} - 6 = 1042$ degrees of 
freedom.  We then determine the equivalent statistical significance to which this 
$p$-value would correspond for a Gaussian distribution.  We consider regions of 
parameter space for which this Gaussian-equivalent statistical significance exceeds 
$3\sigma$ to be excluded.  

In surveying the parameter space of our model, two issues arise which require further
comment.  First, when comparing the $\mu(z_j)$ in the context of any particular model to 
the corresponding measured values $\mu_{j}^\mathrm{obs}$, we must account for 
systematic uncertainties in the overall normalization of the theoretical $\mu(z)$ curve relative 
to this set of measured values.  Indeed, this relative normalization depends both on the value 
of $H_{\mathrm{now}}$ and on the absolute magnitude of the reference population of Type~Ia 
supernovae 
against which the $\mu_{j}^\mathrm{obs}$ are calibrated~\cite{Tripp}, both of which involve
non-negligible uncertainties.  It is not our aim in this paper to perform an analysis of these 
uncertainties or to assess the degree of tension which exists between observational data and 
the predictions of the standard cosmology, but rather to constrain deviations from the 
standard cosmology which result from replacing the stable dark-matter candidate with a DDM 
ensemble on the basis of Type~Ia supernova data alone.  Thus, in our analysis, we shall adopt a 
conservative approach to constraining these deviations in which we adjust the $\mu_j^{\rm obs}$ 
by an overall additive constant chosen such that the goodness-of-fit statistic $\chi^2$ defined 
in Eq.~(\ref{eq:ChiSq}) is minimized for a stable dark-matter candidate in the $\Lambda$CDM 
cosmology.  Possible alternative approaches in which additional cosmological parameters 
are also allowed to vary will be discussed in Sect.~\ref{sec:Conclusions}.~  
We note that as a result of our taking this conservative approach, the bounds we 
obtain on the lifetime of a single unstable particle species which decays to two massless daughter
particles are slightly weaker than those obtained in Ref.~\cite{BlackadderKoushiappas1}.
    
The second issue that we must address concerns our initial conditions for the ensemble. 
Planck data place stringent constraints on the abundances of both dark matter and
dark radiation at $t = \tLS$.  Deviations in the present-day dark matter abundance
$\OmegaDM(\tnow) h^2 \approx 0.120$ inferred from CMB data are constrained at the percent 
level~\cite{Planck}, implying a similar bound on deviations from the dark-matter abundance 
at $t = \tLS$.  The corresponding constraint on the abundance of dark radiation is typically 
phrased as a bound on the net additional contribution $\Delta N_{\mathrm{eff}}$ to the 
effective number of neutrino species $N_{\mathrm{eff}}$ from particles other than SM 
neutrinos.  The current bound $\Delta N_{\mathrm{eff}} \lesssim 0.28$~\cite{Planck}
implies a constraint 
\begin{equation}
  \frac{\Omega_\psi(\tLS)}{\Omega_\gamma(\tLS)} ~\lesssim~ 0.15
  \label{eq:DarkRadBound}
\end{equation}     
on the abundance of the dark-radiation field $\psi$ within our DDM model at the $95\%$ 
confidence level, where $\Omega_\gamma(\tLS)$ denotes the abundance of photons at $t=\tLS$.  
Taken together, these constraints imply that the cosmology of our DDM model should not 
differ significantly from that of a $\Lambda$CDM universe at $t \lesssim \tLS$.

The early decays of the $\chi_n$ --- and especially those with lifetimes in the
regime $\tau_n \lesssim \tLS$ --- can lead to a significant reduction in the total dark-matter
abundance at last scattering and generate a significant abundance for dark radiation by $t=\tLS$.  
We must therefore ensure that the collective effect of these early decays on the cosmology of 
our DDM scenario at times $t \lesssim \tLS$ is negligible.  In doing so, we proceed as follows.
We begin by defining the extrapolated abundance $\widetilde{\Omega}_n(t)$ of $\chi_n$ at 
time $t$, which represents the abundance that this ensemble constituent {\it would have had}\/
at time $t$ if it were absolutely stable.  We fix the initial value $\Omega_0(\tprod)$ by 
demanding that the total extrapolated abundance 
$\widetilde{\Omega}_{\mathrm{tot}}(t) \equiv \widetilde{\Omega}_n(t)$ of the ensemble
is equal to the central value for $\OmegaDM(\tLS)$ inferred from Planck data.  We 
then calculate the actual abundances $\Omegatot(\tLS)$ and $\Omega_\psi(\tLS)$ 
accounting for the effect of $\chi_n$ decay.  In doing so, we approximate the universe as 
radiation-dominated, with $a \propto t^{1/2}$, prior to the time $t_{\mathrm{MRE}}$ 
of matter-radiation equality, and as matter-dominated, with $a \propto t^{2/3}$, 
for $t_{\mathrm{MRE}} < t < \tLS$.  For any given choice of model parameters, 
we define\footnote{We have chosen the Korean word \begin{CJK}{UTF8}{mj}무\end{CJK}, 
pronounced ``mu'' and meaning ``void'' or ``empty,'' as our notation for this parameter, 
since its purpose is to ensure that the universe is essentially devoid or empty of dark 
radiation at times $t \lesssim \tLS$.} a small cutoff parameter 
\begin{CJK}{UTF8}{mj}무\end{CJK} and then impose a constraint
\begin{equation}
  1 - \frac{\Omegatot(\tLS)}{\widetilde{\Omega}_{\mathrm{tot}}(\tLS)} 
    ~\leq~ \mbox{\begin{CJK}{UTF8}{mj}무\end{CJK}}
  \label{eq:MuTest}
\end{equation}     
on the portion of the overall dark-matter abundance that has be depleted by decays
prior to last scattering.  Any ensemble which does not satisfy this criterion is considered 
to be inconsistent with our initial conditions and therefore excluded.  Given the aforementioned 
constraint on the dark-matter abundance, we take \begin{CJK}{UTF8}{mj}무\end{CJK}$\, = 0.01$. 
We note that for this value of \begin{CJK}{UTF8}{mj}무\end{CJK}\,, the constraint in
Eq.~(\ref{eq:MuTest}) is always more stringent than the corresponding constraint 
on $\Omega_\psi(\tLS)$ from Eq.~(\ref{eq:DarkRadBound}).


\section{Results\label{sec:Results}}

\begin{figure}
  \includegraphics[width=0.95\linewidth]{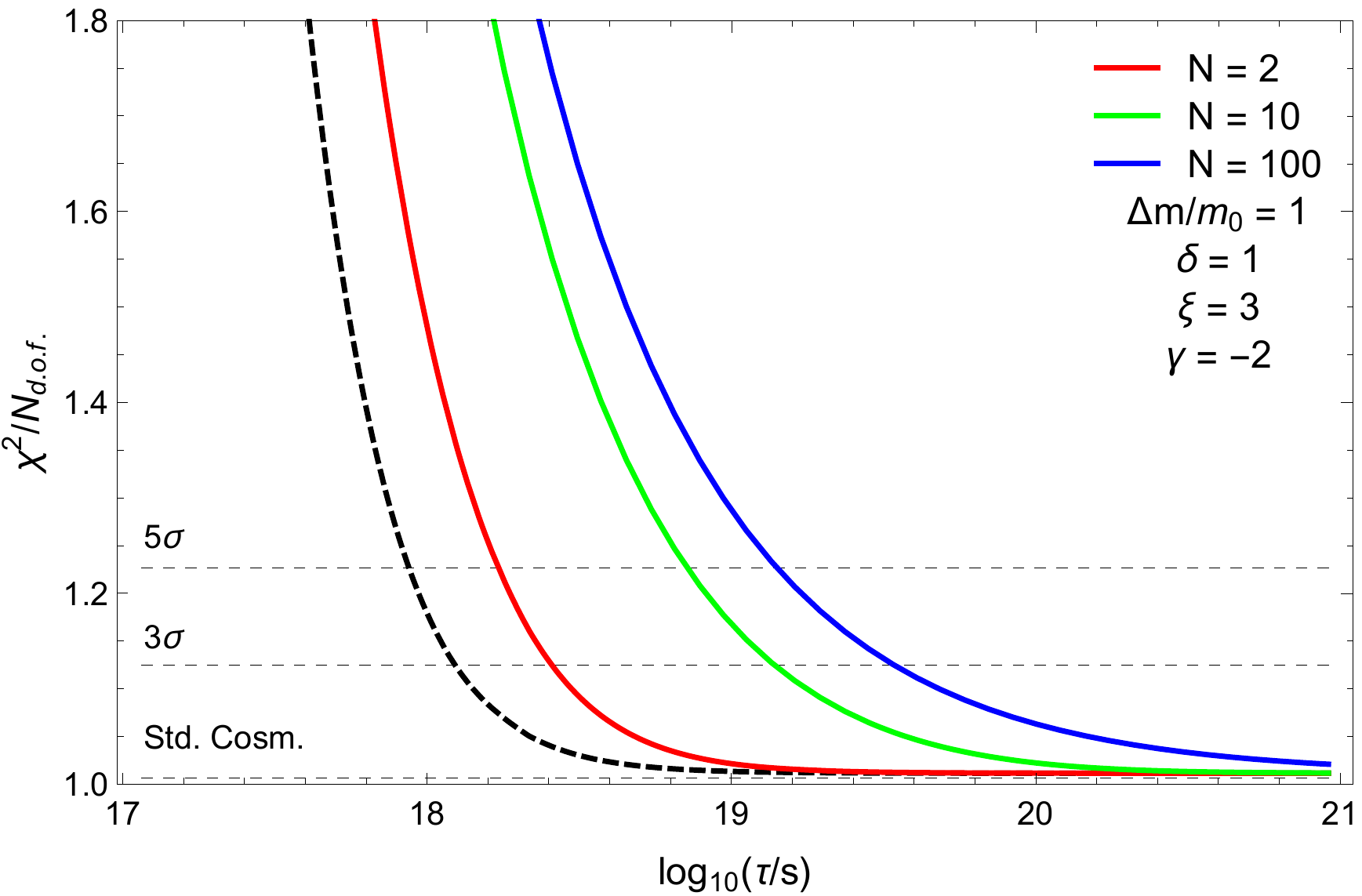}\\
  \includegraphics[width=0.95\linewidth]{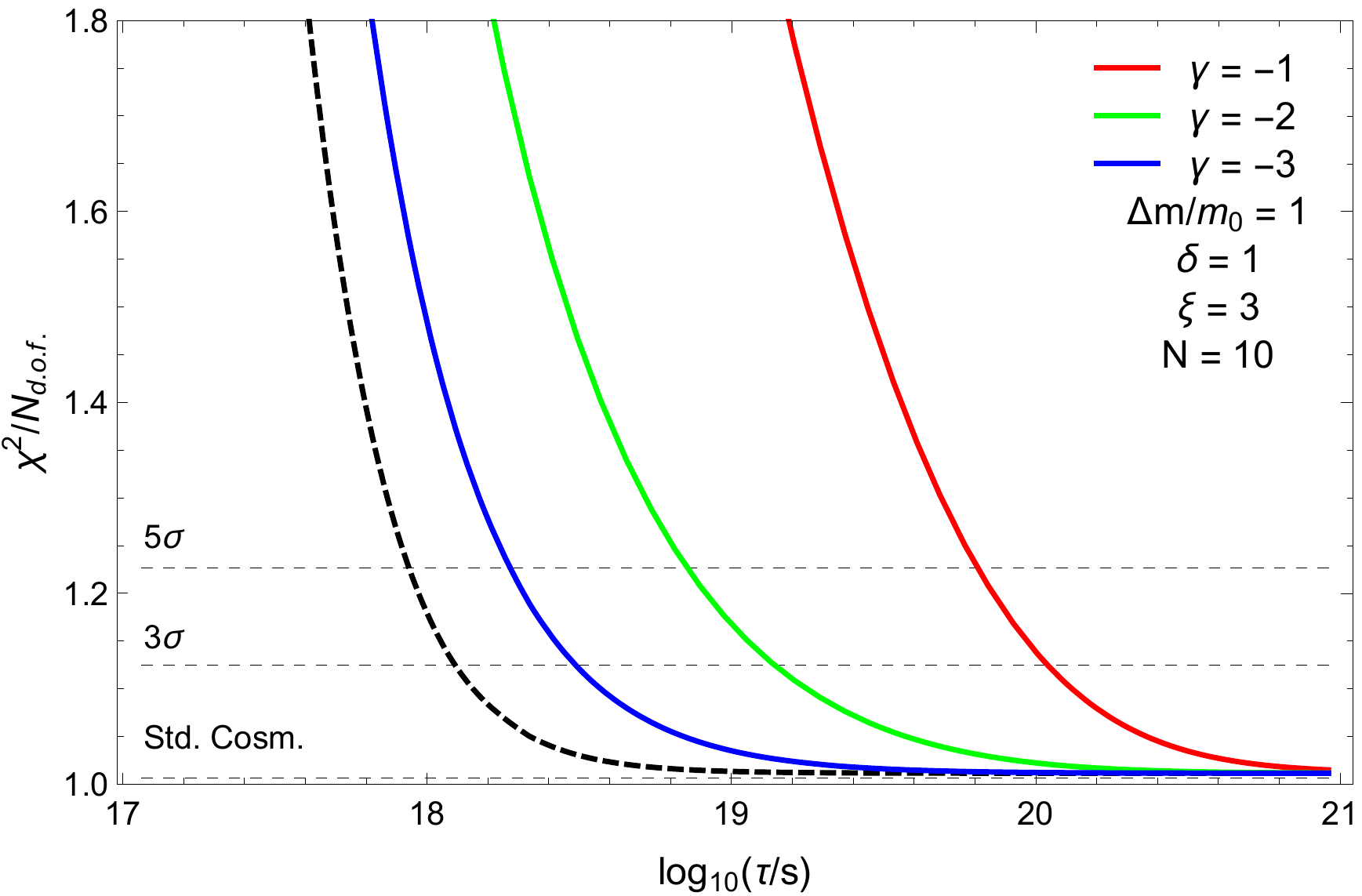}
\caption{The goodness of fit $\chi^2/\Ndof$ between the distance-modulus function 
  $\mu(z)$ obtained for a particular set of 
  DDM model parameters and the data in the Pantheon sample, plotted as functions of the 
  lifetime $\tau_0$ of the lightest particle in the DDM ensemble.  The curves shown in the 
  top panel correspond to different values of $N$ for fixed $\gamma=-2$,
  $\Delta m/m_0 = 1$, and $\xi=3$.  The curves shown in the bottom panel correspond to 
  different values of $\gamma$ for fixed $N = 10$ and the same values of $\Delta m/m_0$ and
  $\xi$.  In each panel, the corresponding curve for a single dark-matter particle species with 
  lifetime $\tau_0$ is indicated by the black dashed curve.  For reference, within each panel 
  we have also included dashed lines showing the values of $\chi^2/\Ndof$ which 
  correspond to a discrepancy between the theoretical $\mu(z)$ function and the Pantheon data 
  at the $3\sigma$ and $5\sigma$ significance levels, along with another dashed line indicating 
  the value of $\chi^2/\Ndof$ obtained for a stable dark-matter candidate in the standard 
  cosmology.         
  \label{fig:ChisQCurves}}
\end{figure}

We begin the discussion of our results by examining how the goodness of fit
between the theoretical distance-modulus function $\mu(z)$ and the Pantheon data varies 
across the parameter space of our DDM model.  In Fig.~\ref{fig:ChisQCurves}, we 
display curves showing the value of $\chi^2/\Ndof$ as a function of 
$\tau_0 \equiv \Gamma_0^{-1}$ for different choices of the model parameters $N$ (top panel) 
and $\gamma$ (bottom panel) which essentially control the distribution of 
$\Omegatot(\tLS)$ across the ensemble.  The results shown in the top panel correspond to 
the choices $\gamma = -2$, $\Delta m/m_0 = 1$, and $\xi=3$; the results
shown in the bottom panel correspond to the choices $N = 10$ and the same values of
$\Delta m/m_0$ and $\xi$.  In each panel, for reference, we have also included a (dashed black) 
curve showing $\chi^2/\Ndof$ for a single dark-matter particle species with 
lifetime $\tau_0$.  The lower dashed horizontal line in each panel indicates the value of 
$\chi^2/\Ndof$ obtained for a stable dark-matter particle in the standard cosmology,
while the other dashed horizontal lines indicate the values of $\chi^2/\Ndof$ for which 
the corresponding $p$-values would be associated with the statistical significances 
$3\sigma$ and $5\sigma$ for a Gaussian distribution. 
  
The results shown in the top panel of Fig.~\ref{fig:ChisQCurves} illustrate the impact 
on $\chi^2/\Ndof$ of introducing additional, unstable states into the ensemble.  The
results shown indicate that the value of $\chi^2/\Ndof$ is quite sensitive to the value
of $N$ in the regime in which $N$ is small, but becomes less sensitive as $N$ increases.  
By contrast, the results shown in the bottom panel illustrate that $\chi^2/\Ndof$ becomes
increasingly sensitive to the value of $\gamma$ as $\gamma$ increases. 

\begin{figure*}[t!]
  \includegraphics[width=0.27\linewidth]{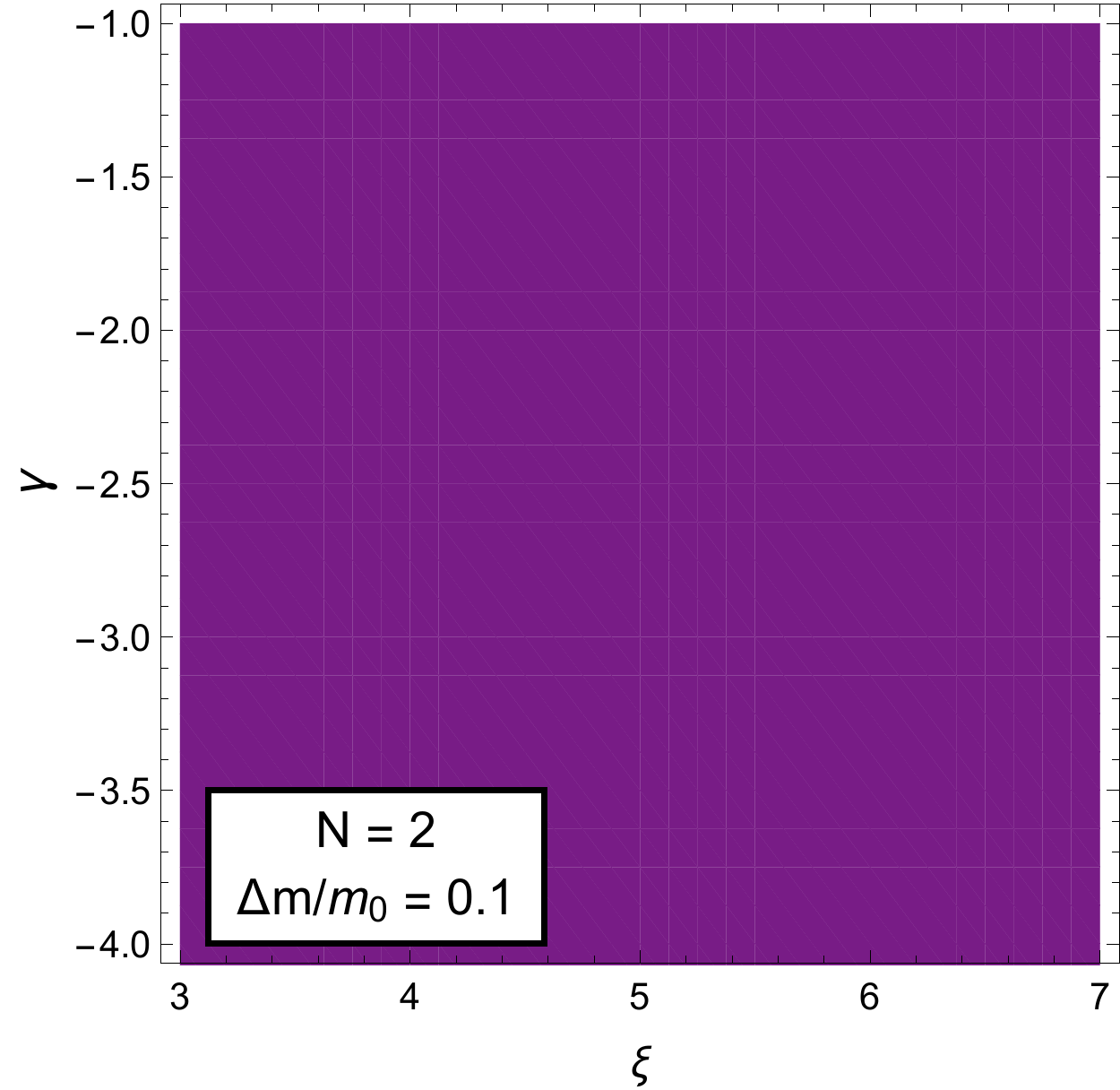}
  \includegraphics[width=0.27\linewidth]{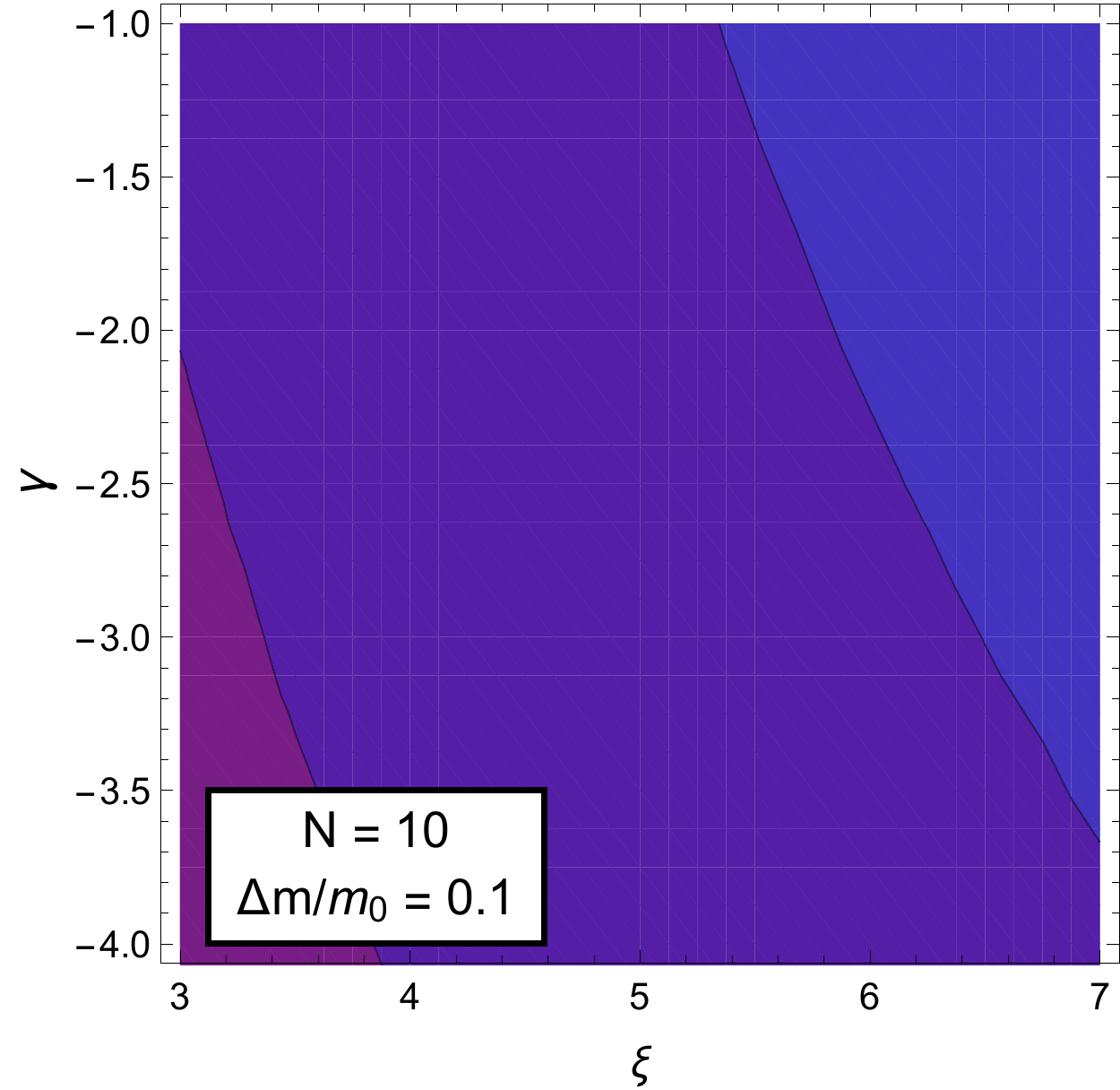}
  \includegraphics[width=0.27\linewidth]{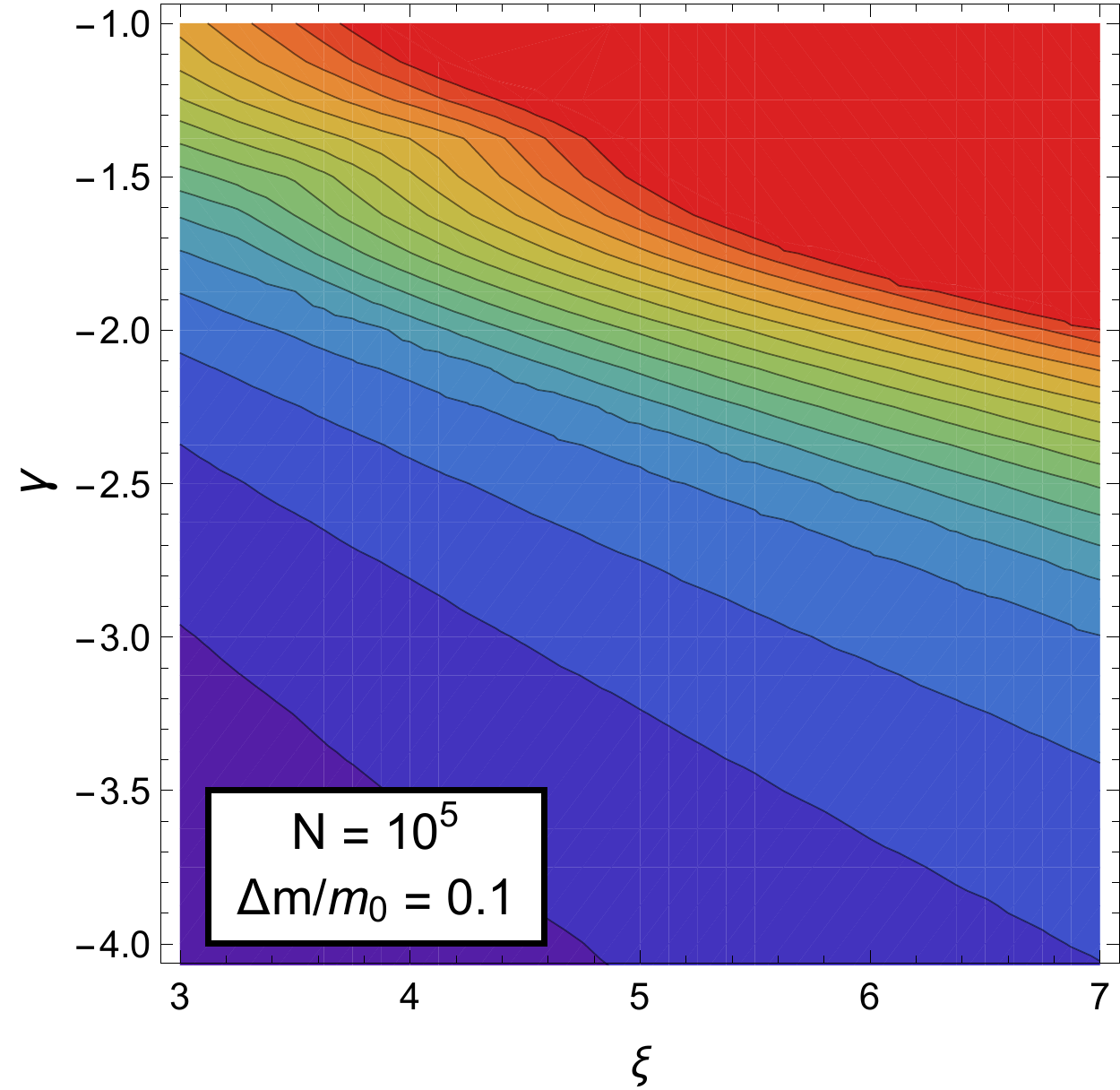}\\
  \includegraphics[width=0.27\linewidth]{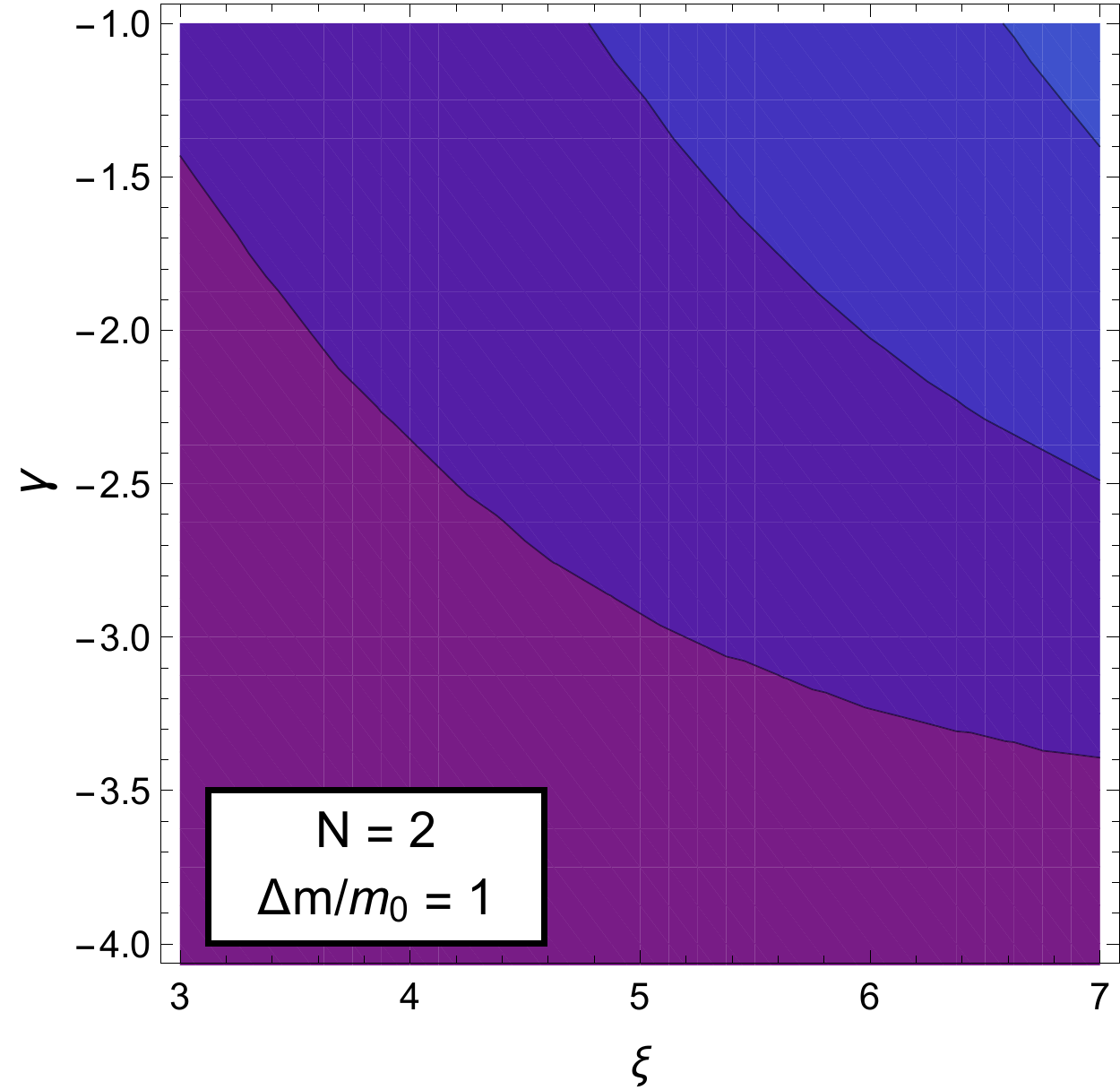}
  \includegraphics[width=0.27\linewidth]{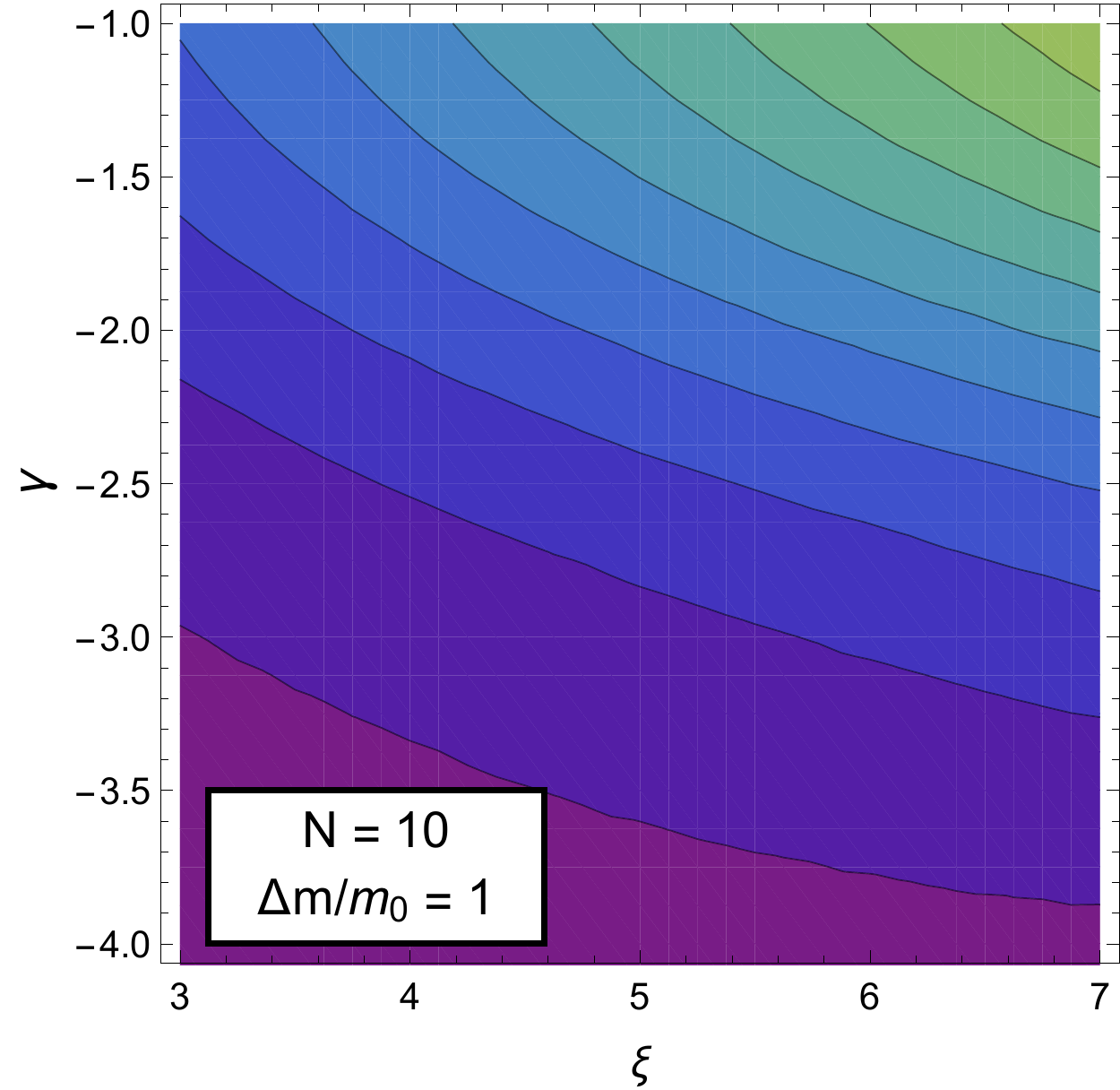} 
  \includegraphics[width=0.27\linewidth]{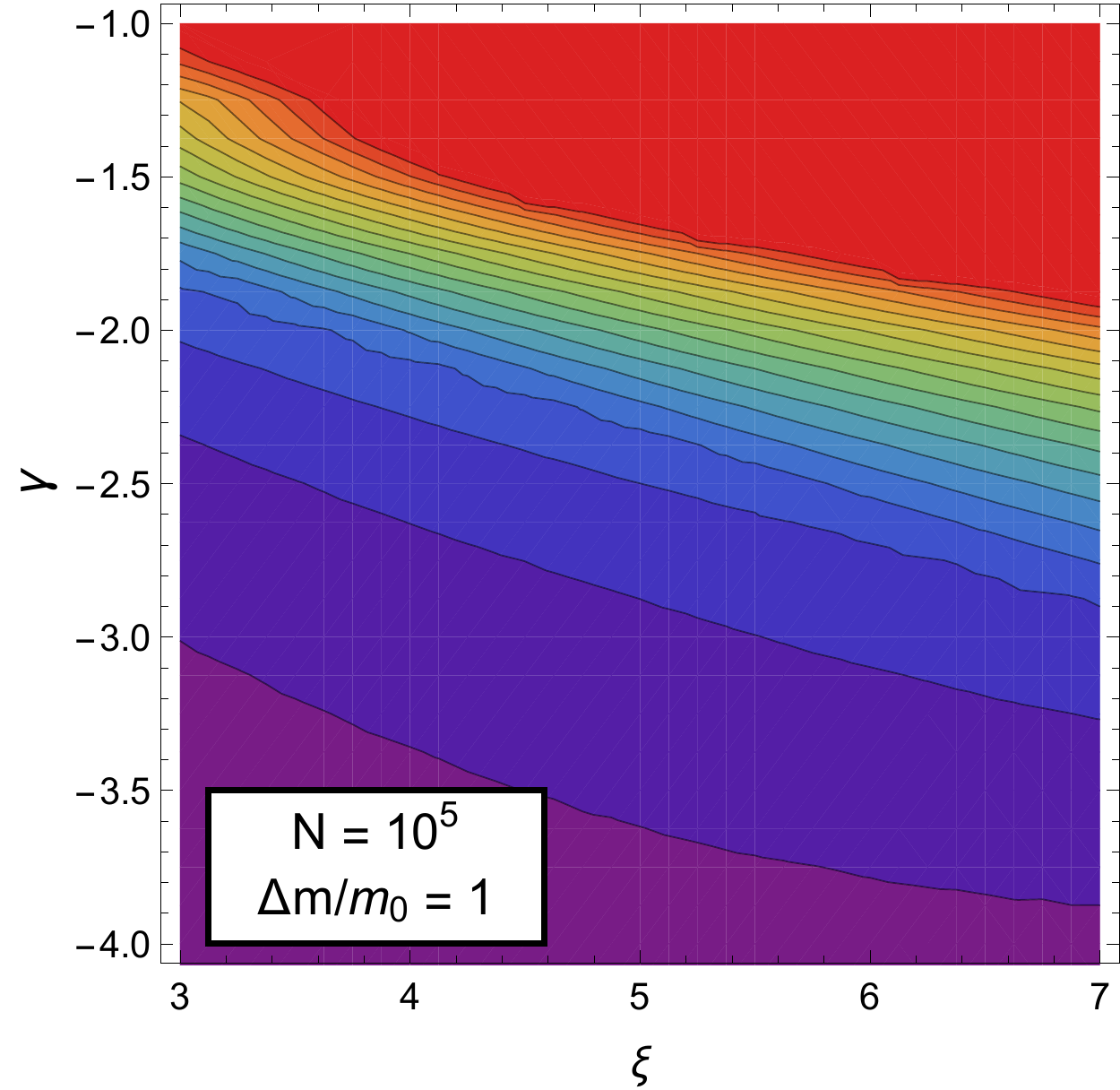}\\
  \includegraphics[width=0.27\linewidth]{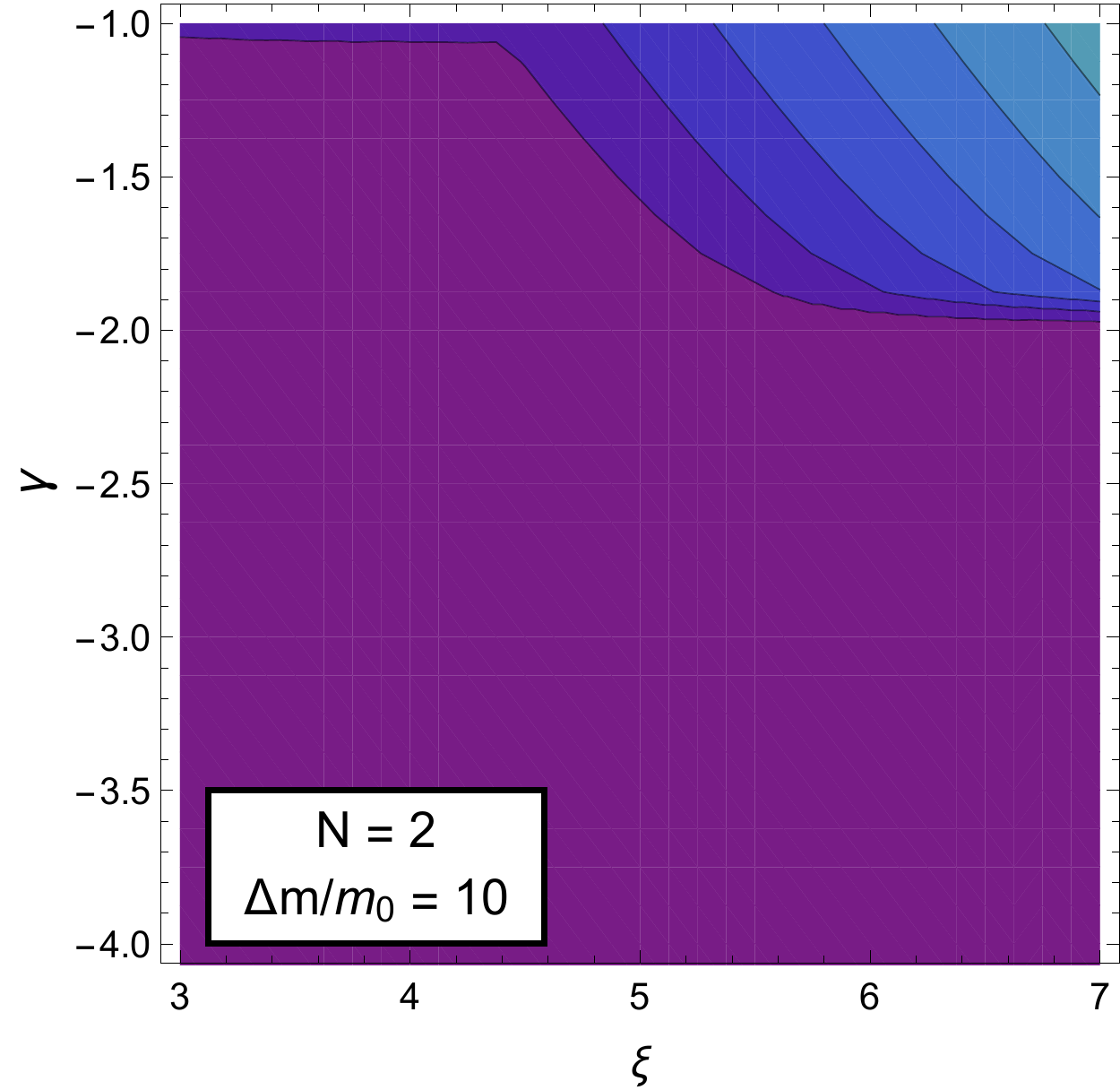}
  \includegraphics[width=0.27\linewidth]{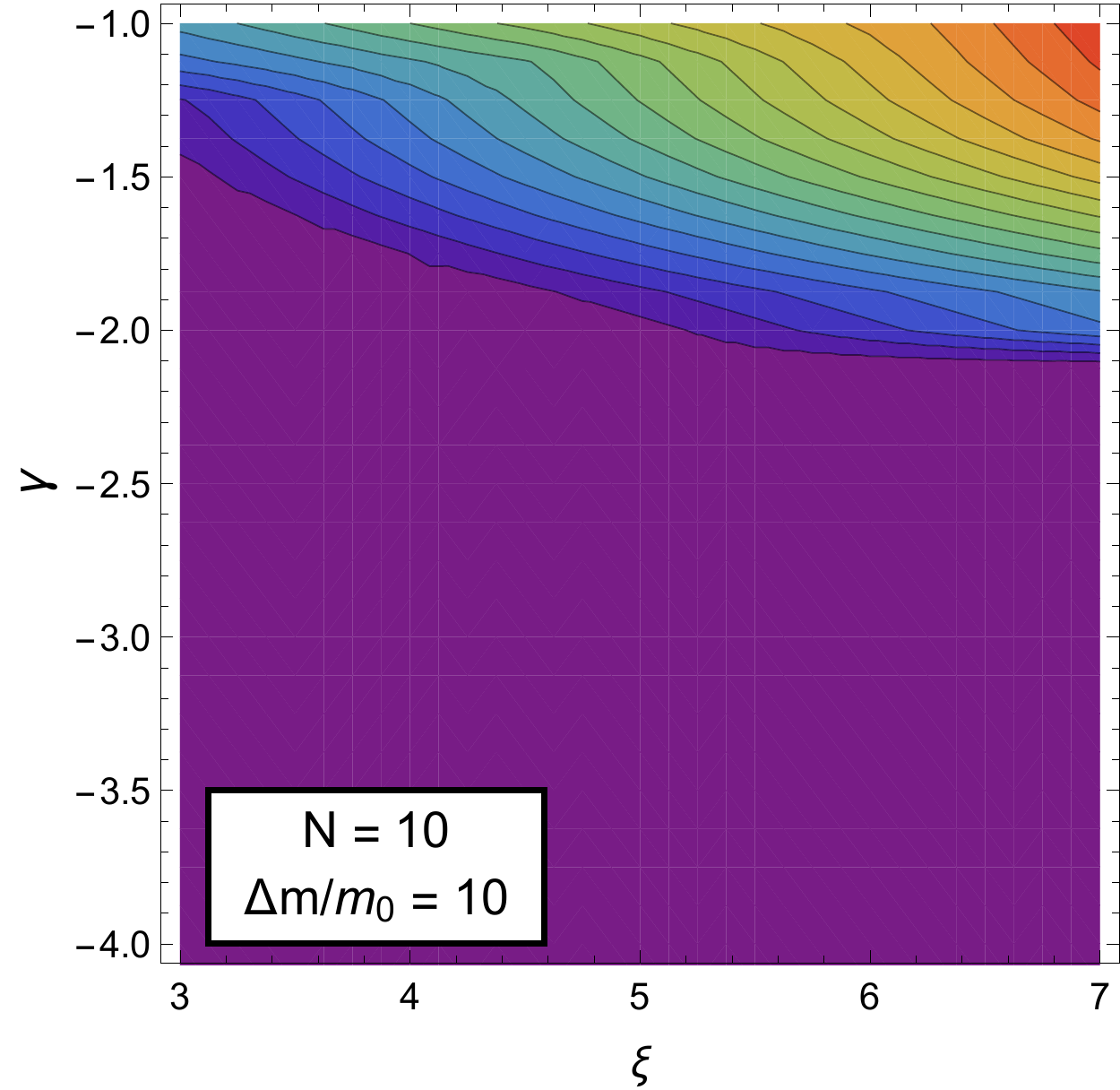}
  \includegraphics[width=0.27\linewidth]{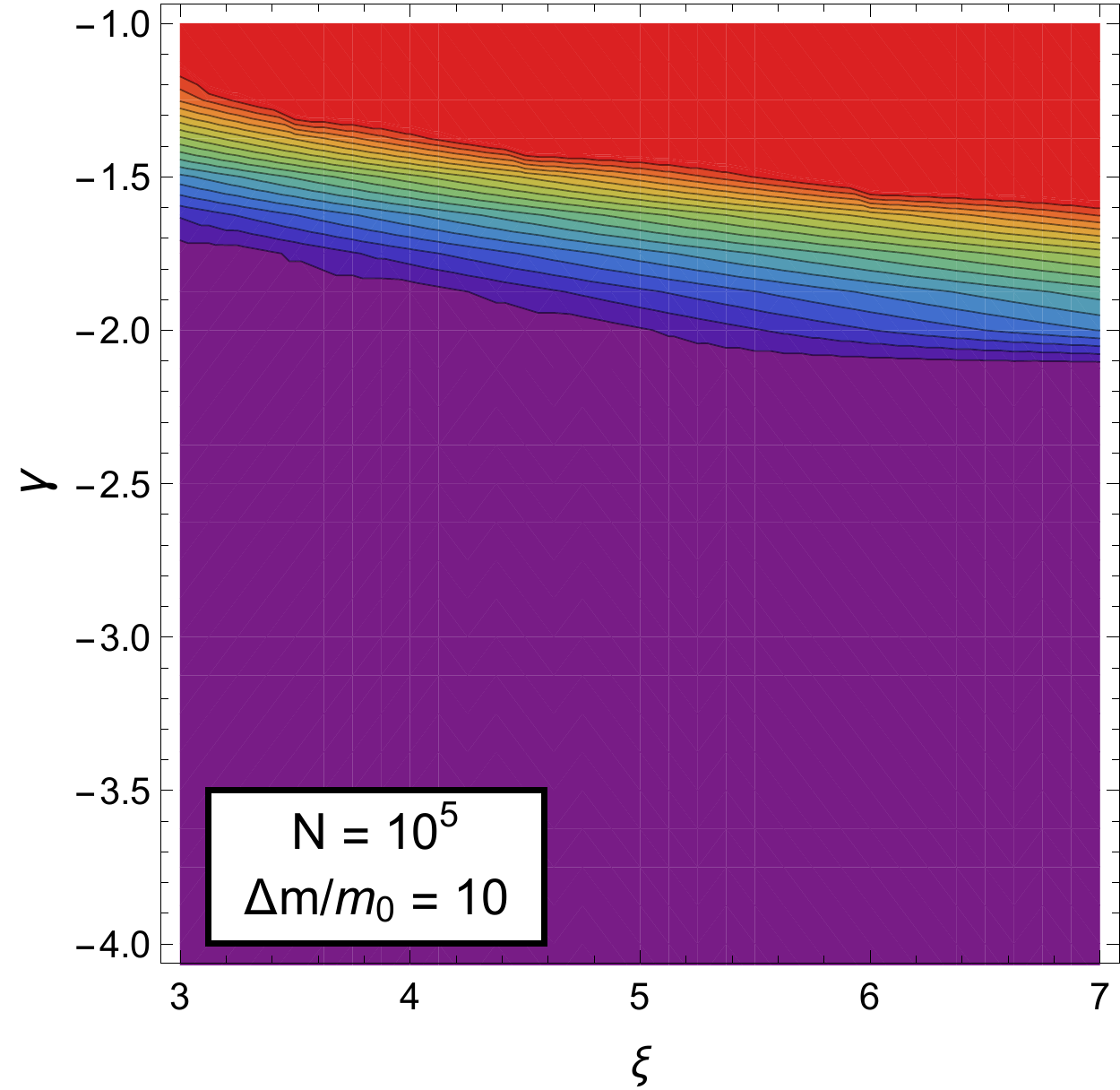}\\
  \includegraphics[width=0.65\linewidth]{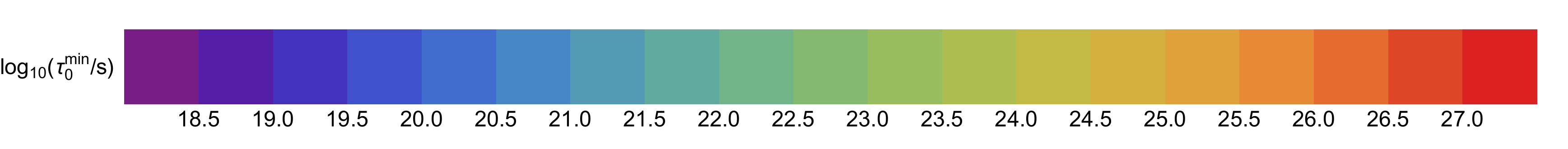}
\caption{Constraints on the parameter space of our DDM model from supernova data.  
  In each panel, the contour plot shows the $3\sigma$ lower bound $\tau_0^{\rm min}$ on 
  the lifetime of the lightest particle in the ensemble for a particular choice of $N$ and 
  $\Delta m/m_0$.  The results displayed in the different columns of the figure from left to 
  right respectively correspond to the parameter choices choices $N = \{2, 10, 10^5\}$. Likewise, 
  the results displayed in the different rows of the figure from top to bottom respectively 
  correspond to the parameter choices $\Delta m/m_0 = \{0.1, 1 , 10\}$.
  We see that our constraints are generally most severe for ensembles with intermediate values 
  of $\Delta m/m_0 \sim \mathcal{O}(1 - 10)$ and large values of $\xi$ and $\gamma$. 
  \label{fig:ExclusionPanels}}
\end{figure*}

In Fig.~\ref{fig:ExclusionPanels}, we display the constraints on the parameter
space of our DDM model from the Pantheon data sample.  The contour plot in each panel 
of the figure shows the $3\sigma$ lower bound $\tau_0^{\rm min}$ on $\tau_0$ 
within the $(\xi,\gamma)$-plane for a particular choice of $N$ and $\Delta m/m_0$.
The results displayed in the different columns of the figure from left to right 
respectively correspond to the parameter choices $N = \{2, 10, 10^5\}$.  Likewise, the 
results displayed in the different rows of the figure from top to bottom respectively 
correspond to the parameter choices $\Delta m/m_0 = \{0.1, 1 , 10\}$.    
 
Interpreting the results shown in Fig.~\ref{fig:ExclusionPanels}, we first note 
that within each individual panel of the figure, $\tau_0^{\rm min}$ generically 
increases as both $\xi$ and $\gamma$ are increased.  This is to be expected: increasing 
$\gamma$ redistributes a larger fraction of $\Omegatot(\tLS)$ to the heavier, more 
unstable modes in the ensemble, while increasing $\xi$ decreases the lifetimes of these 
heavier modes.  Likewise, comparing the results across different panels of the figure, we 
see that $\tau_0^{\rm min}$ generically increases as we move from left to right across
the panels within any given row of the figure --- \ie, as we increase $N$ while holding
$\Delta m/m_0$ fixed.  Indeed, this behavior accords with the results shown in the top 
panel of Fig.~\ref{fig:ChisQCurves}.  

However, the way in which $\tau_0^{\rm max}$ changes as we move from top to bottom along 
the panels within a given column of the figure --- \ie, as we increase $\Delta m/m_0$ while 
holding $N$ fixed --- is far less straightforward and depends non-trivially on the value 
of $N$.  When $\Delta m/m_0$ is taken to be sufficiently small for any finite value of $N$,
the lifetimes of all of the $\chi_n$ with $n > 0$ are comparable to $\tau_0$.  Thus, in this
regime, the ensemble effectively behaves like a single unstable particle with lifetime $\tau_0$
in terms of its effect on the expansion history of the universe.  Moreover, in this regime, 
$\tau_0^{\rm max}$ is fairly insensitive to the values of $\gamma$ and $\xi$.  The results
shown in the top left panel of the figure, which correspond to the parameter choices $N=2$ 
and $\Delta m/m_0 = 0.1$, are representative of this regime.  The larger $N$ is, however, the 
greater the range of masses present within the ensemble and the smaller $\Delta m/m_0$ must 
therefore be in order for the ensemble to remain in this regime.  Indeed, even for $N = 10$, we 
see that $\Delta m/m_0$ is sufficiently large that $\tau_0^{\rm max}$ exhibits a non-trivial
dependence on $\xi$ and $\gamma$.     

On the other hand, for intermediate values of $\Delta m/m_0 \sim \mathcal{O}(1 - 10)$ 
we observe that the results for $\tau_0^{\rm max}$ can differ considerably from the 
corresponding constraints on a single decaying particle.  Indeed, for such values of 
$\Delta m/m_0$, the $\chi_n$ with $n>0$ not only collectively represent a non-negligible 
fraction of $\Omegatot(\tLS)$ within the region shown in each panel, but also exhibit a 
broad range of lifetimes.  Thus, it is for DDM ensembles with 
$\Delta m/m_0 \sim \mathcal{O}(1 - 10)$ that supernova data are generally the most 
constraining.

It is also worth noting that it is within this intermediate-$\Delta m/m_0$ regime --- and
especially when $N$ is small --- that the constraint in Eq.~(\ref{eq:MuTest}) has a significant 
impact on the value of $\tau_0^{\rm max}$.  In the panel of Fig.~\ref{fig:ExclusionPanels}
corresponding to $N = 2$ and $\Delta m/m_0 = 10$, for example, this constraint plays a
crucial role in establishing the lifetime bound obtained within the region of the 
$(\xi,\gamma)$-plane wherein $\xi$ is large and $\gamma \lesssim -2$.  This can be
understood as follows.  In the regime in which $\Delta m/m_0 \gtrsim 1$, an extreme
value of $\xi$ is not required in order to achieve a significant difference between the
lifetimes of the lightest two constituents in the DDM ensemble.  
Indeed, for any fixed value for $\tau_0$, the lifetime $\tau_1$ of $\chi_1$ decreases as 
$\xi$ increases and eventually becomes comparable to $\tLS$.  This implies that a significant 
fraction of the abundance $\widetilde{\Omega}_1(\tLS)$ which this ensemble constituent would 
have had at last scattering had it been stable is instead converted to dark radiation prior 
to $\tLS$.  For $\gamma \geq -2$, this $\widetilde{\Omega}_1(\tLS)$ represents a
sufficiently large fraction of $\widetilde{\Omega}_{\mathrm{tot}}(\tLS)$ that a 
sizable value of $\tau_0$ is required in order not to violate the 
the constraint in Eq.~(\ref{eq:MuTest}).  This constraint also imposes a similar lower bound 
on $\tau_0$ in the other panels of Fig.~\ref{fig:ExclusionPanels} for which $\Delta m/m_0 = 10$, 
but this bound is superseded by the $3\sigma$ lower bound on $\tau_0$ from the Pantheon data 
throughout most of the same region of the $(\xi,\gamma)$-plane for both $N = 10$ and $N = 10^5$.  

The $\tau_0^{\rm max}$ contours obtained for even larger values of $\Delta m/m_0$ 
follow the general trends exhibited in Fig.~\ref{fig:ExclusionPanels}. 
Indeed, when $\Delta m/m_0 \gg 1$, the vast majority of $\Omegatot(\tLS)$
is carried by $\chi_0$ unless the value of $\gamma$ is quite large.  As a result, the bound
on $\tau_0$ typically becomes weaker with increasing $\Delta m/m_0$ within this regime.
Thus, if one were to plot contours of $\tau_0^{\rm max}$ for $\Delta m/m_0 \gg 1$ similar
to those shown in Fig.~\ref{fig:ExclusionPanels}, one would find that the value of 
$\tau_0^{\rm max}$ would not significantly differ from the lower bound on the lifetime 
of a single unstable particle which decays to dark radiation throughout most of the
same region of the $(\xi,\gamma)$-plane, regardless of the value of $N$.

Given the results in Fig.~\ref{fig:ExclusionPanels}, it is likewise straightforward to 
infer the behavior of the corresponding $\tau_0^{\rm max}$ contours for even larger values 
of $N$.  As illustrated in Fig.~\ref{fig:ChisQCurves}, $\tau_0^{\rm max}$ becomes largely 
insensitive to $N$ in the regime in which $N$ is large.  Indeed, provided that 
$\gamma < -1$ and that the sum over $\Omega_n(\tprod)$ converges in the $N\rightarrow \infty$ 
limit, one finds that $\tau_0^{\rm max}$ approaches a finite asymptotic value this limit.  
Thus, for any particular choice of the remaining model parameters, the value of 
$\tau_0^{\rm max}$ obtained for sufficiently large, finite $N$ is effectively 
equal to this asymptotic value.  Throughout most the region of the $(\xi,\gamma)$-plane 
shown in the panels of Fig.~\ref{fig:ExclusionPanels}, the number $N = 10^5$ turns 
out to be sufficiently large that the value of 
$\tau_0^{\rm max}$ obtained for this choice of $N$ lies within this asymptotic regime.  
Indeed, only when  $\gamma$ approaches the value 
$\gamma = -1$ at which the sum over $\Omega_n(\tprod)$ formally diverges do the results for 
$\tau_0^{\rm max}$ begin to deviate significantly from those obtained in the 
$N \rightarrow \infty$ limit. 

The results shown in Fig.~\ref{fig:ExclusionPanels} demonstrate that meaningful bounds on the
parameter space of DDM ensembles can be derived from constraints on the expansion history 
of the universe --- and in particular on the relationship the between redshifts and luminosity
distances of Type~Ia supernovae --- even in situations in which the decays of the ensemble 
constituents decay entirely to other light states within a hidden sector.  These bounds turn
out to be the most constraining for ensembles with intermediate values of 
$\Delta m/m_0 \sim \mathcal{O}(1 - 10)$ and large values of $\xi$ and $\gamma$.



\section{Conclusions\label{sec:Conclusions}}


In this paper, we have considered the constraints on DDM ensembles 
whose constituent particles decay primarily to other, lighter particles within the 
dark sector by analyzing the constraints on such ensembles which arise
from the effects of these decays on the expansion history of the universe.  In 
particular, we have derived constraints on the parameter space of such ensembles 
from the relationship between the observed redshifts and luminosity distances 
of Type~Ia supernovae within the Pantheon data sample.

Several comments are in order.  First, we note that a variety of other 
considerations can also be used to constrain the decays of dark-sector particles to other
states within the dark sector.  For example, baryon acoustic oscillations and the 
properties of the CMB both provide information about the expansion history of the 
universe.  These considerations have been used to constrain dark-matter decays within 
single-component dark-matter scenarios~\cite{BlackadderKoushiappas2}, and it would be
interesting to examine the extent to which these complementary probes of the expansion 
history at different redshifts constrain the parameter space of DDM ensembles
as well.  In addition, decays within the dark sector can also give rise to characteristic 
features within the matter power spectrum which can yield information about the structure 
of the decaying ensemble~\cite{Archaeology}. 

In addition, in this paper we have focused on
ensembles in which each constituent $\chi_n$ decays exclusively to dark radiation.  As
discussed in Sect.~\ref{sec:DDMEnsemble}, the corresponding constraints on ensembles
in which the $\chi_n$ can also decay into final states involving other, lighter ensemble 
constituents are always less stringent, given that decays directly to dark radiation 
represent the most efficient conversion possible of mass energy into kinetic energy. 
It would nevertheless be interesting to investigate the supernova constraints on 
ensembles in which intra-ensemble decays play a significant role in the decay phenomenology 
of the $\chi_n$.          

Finally, as discussed in Sect.~\ref{sec:SNData}, in this paper we have focused 
on modifications of the standard cosmology in which the stable dark-matter candidate 
is replaced with a DDM ensemble, but in which no further modifications are made. 
Moreover, we have assumed that the values of the relevant nuisance parameters are such 
that the $d_L(z)$ function obtained for the standard cosmology provides a good fit to 
the Pantheon data.  These include several nuisance parameters involved in determining 
the $\mu_j^{\rm obs}$ for the supernovae in the Pantheon sample.
While this is common practice~\cite{Fulvio1,BlackadderKoushiappas1}, we note that
the assessment of the statistical likelihood for any cosmological model
can be improved by simultaneously fitting the values of these nuisance parameters 
along with with the values of the parameters which characterize that 
model~\cite{Fulvio2,Fulvio3}.  While the complexity of our DDM model renders such an
analysis impractical for a survey of the sort we have undertaken in this paper,  
a study along these lines would be an interesting avenue for future research. 

Along the same lines, while this minimal approach is fruitful for constraining deviations 
from the standard cosmology within the DDM framework, there are compelling reasons 
why it would be interesting to consider a more general study in which other cosmological 
parameters are allowed to vary.
For example, a statistically significant tension currently exists between the value 
of $H_{\mathrm{now}}$ obtained from local probes of the cosmic expansion rate 
(including not only Type~Ia supernova 
data~\cite{RiessHubbleTension1,RiessHubbleTension2,RiessHubbleTension3,Colgain}, 
but also lensing time-delay experiments~\cite{H0LiCOW1,H0LiCOW2}) and the value 
inferred from CMB
data in the context of a $\Lambda$CDM cosmology~\cite{EfstathiouHubbleTension,
AddisonHubbleTension,AghanimHubbleTension,AylorHubbleTension}.  
Dark-matter decays between $\tLS$ and $\tnow$ have 
been posited as one possible~\cite{EnqvistDMDecayHT,LuisDMDecayHT,BuchDMDecayHT,
BringmannDMDecayHT,PandeyDMDecayHT,SavvasHubbleTension} way of alleviating these tensions.  

While it has not been our aim in this paper to address the Hubble tension, it likely that DDM 
scenarios of the sort can serve to alleviate this tension.  In order to understand why this 
occurs, we begin by noting that in addition to constraining the energy density of dark matter 
at last scattering, CMB data also tightly constrain the angular horizon size $\theta_s$, which 
in a flat universe may be written as
\begin{equation}
  \theta_s ~=~ \frac{(1+\zLS)r_{\rm LS}}{d_L(\zLS)}~,
\end{equation}
where $r_{\rm LS}$ is the sound horizon and where $d_L(\zLS)$ is the luminosity
distance of the last-scattering surface.  Since the sound horizon is determined by the
state of the universe prior to last scattering, the value of $r_{\rm LS}$ obtained in our 
DDM scenario does not differ appreciably from that obtained in the standard cosmology for
the same choice of cosmological parameters.  By contrast, $d_L(\zLS)$ depends on the state 
of the universe at all redshifts $0 < z < \zLS$.  Thus, its value for a given choice of  
DDM model parameters in general differs --- potentially significantly --- from that obtained 
in the standard cosmology.

In our DDM scenario, decays of the $\chi_n$ between $\tLS$ and $\tnow$
transfer energy density from dark matter to dark radiation.  Since $\rho_\psi$ 
decreases more rapidly as a result of cosmic expansion than does $\rho_\chi$, the expansion 
rate of the universe is lower in our DDM scenario at low redshifts than it would have been in a 
$\Lambda$CDM scenario with the same value $H(\zLS)$ of the Hubble parameter at last scattering.  
However, a consistently lower value of $H(z)$ at late times results in a larger value for 
$d_L(\zLS)$.  This in turn results in a smaller value for $\theta_s$.  Thus, in order to obtain 
a value of $\theta_s$ which accords with Planck data in our DDM scenario, we need to compensate 
for this decrease by increasing the dark-energy density $\rho_\Lambda$, which in turn increases 
$H(\zLS)$.  The larger $\rho_\Lambda$ implies that dark energy will begin to dominate the 
universe at a slightly earlier time than it otherwise would in the standard cosmology, and 
consequently yields a larger value of $H_{\rm now}$.  In this sense, our DDM scenario modifies 
the cosmic expansion rate in the right direction for addressing the Hubble tension.

We note that this basic mechanism through which DDM can alleviate the Hubble tension
is the same as that which underlies other scenarios for alleviating this
tension through decaying dark matter.  The primary difference, however, is that the DDM 
framework allows the conversion of dark matter to dark radiation to occur more smoothly 
over a longer timescale.  Of course, more quantitative statements concerning the 
degree to which a DDM ensemble can alleviate the Hubble tension would require  
a more detailed study including an analysis of how the decays of the $\chi_n$ would 
collectively impact the properties of the CMB, the matter power spectrum, and other 
cosmological observables.  We leave such a study for future work. 


\begin{acknowledgments}


We are happy to thank S.~Koushiappas for discussions.  AD also wishes to thank 
the EXCEL Scholars Program for Undergraduate Research at Lafayette College, which 
helped to facilitate this research.  The research activities of AD and BT are supported 
in part by the National Science Foundation under Grant PHY-1720430.  The research 
activities of KRD are supported in part by the Department of Energy under 
Grant DE-FG02-13ER41976 (DE-SC0009913) and by the National Science Foundation through 
its employee IR/D program.  The opinions and conclusions expressed herein are those of 
the authors, and do not represent any funding agencies. 
 
\end{acknowledgments} 


\end{document}